\documentclass{elsart}
\usepackage{epsfig}
\usepackage{amssymb}
\begin{document}
\begin{frontmatter}
\title{Neutrino Propagation Through Matter}
\author{Vadim A. Naumov\thanksref{emVN}}
\address{Laboratory of Theoretical Physics and Physics Department,
         Irkutsk State University, Gagarin boulevard 20, RU-664003
         Irkutsk, Russia \\ and \\
         INFN, Sezione di Firenze, Largo Enrico Fermi 2,
         I-50125 Firenze, Italy}
\author{Lorenzo Perrone\thanksref{emLP}}
\address{Dipartimento di Fisica, Universit\`{a} degli Studi di
         Firenze \\ and \\
         INFN, Sezione di Firenze, Largo Enrico Fermi 2,
         I-50125 Firenze, Italy}
\thanks[emVN]{E-mail: naumov@api.isu.runnet.ru,
                      naumov@fi.infn.it}
\thanks[emLP]{E-mail: perrone@fi.infn.it}
\begin{abstract}
We discuss a simple approach to solve the transport equation for
high-energy neutrinos in media of any thickness.
We present illustrative results obtained with some specific
models for the initial spectra of $\nu_\mu$ and $\overline{\nu}_\mu$
propagating through a normal cold medium.

\end{abstract}
\begin{keyword}
Transport theory; Neutrino interactions \\
\PACS 05.60.+w; 13.15.+g; 14.60.Lm
\end{keyword}
\end{frontmatter}

\newpage
\protect\section{Introduction}\label{sec:Int}

In passing through a medium, neutrinos (and antineutrinos) are
absorbed and lose their energy due to charged and neutral current
interactions.  For a normal \emph{cold} medium, like the earth's or
stellar interior, these are $\nu N$ ($\overline{\nu}N$) and $\nu e$
($\overline{\nu}e$) collisions. In more exotic media, like the relic
black-body neutrino background~\cite{bb} or \emph{hot} galactic halos
filled by massive neutrinos~\cite{halo,GZK}, ultrahigh energy
neutrinos may scatter elastically or be absorbed due to
$\nu\overline{\nu}$ interactions. Owing to the energy loss and the
strong energy dependence of total cross sections for neutrino
interactions, the neutrino ``depth-intensity relation'' (or
penetration coefficient~\cite{Berezinsky86}) does not follow a simple
absorption law and the magnitude of this effect grows with energy and
depth. Such a situation is well known in the muon transport theory
(see e.g. ref.~\cite{Naumov94}). To solve the neutrino transport
equation for moderate depths, the method of successive generations is
workable~\cite{Bugaev95}. But this method becomes inefficient for the
depth in excess of several neutrino interaction lengths. Pertinent
refining of the neutrino transport theory is desirable for many
applications, specifically for studying standard and speculative
neutrino interactions (see e.g. ref.~\cite{exotics} and references
therein),
%%%%%%%%%%%%%%%%%%%%%%%%%%%%%%%%%%%%%%%%%%%%%%%%%%%%%%%%%%%%%%%%%%%%%
%% like direct-channel production of superpartner resonances
%% through R-parity-violating couplings and so on.
%%%%%%%%%%%%%%%%%%%%%%%%%%%%%%%%%%%%%%%%%%%%%%%%%%%%%%%%%%%%%%%%%%%%%
detecting neutrino signals from annihilation of dark matter particles
in the sun and the earth, and for high-energy neutrino astronomy with
future, km$^3$-scale neutrino
telescopes~\cite{Berezinsky90,Learned94}.

The goal of this work is to provide an elementary method for the
precise calculation of the energy spectra of high-energy neutrinos
after their propagation through a medium of \emph{any} thickness. The
problem was considered recently in ref.~\cite{Nicolaidis96} in the
framework of a simplified models for the neutrino cross sections and
initial neutrino spectrum. Our approach does not require
simplifications and is applicable to cross sections (differential and
total) and initial spectra of any form. In sections~\ref{sec:Z} and
\ref{sec:NumRes}, we shall consider only decreasing unbroken initial
spectra most interesting for high-energy neutrino astrophysics.
However, the main idea of the method can be extended also to a
monochromatic spectrum (see appendix~\ref{app:MIS}). This
generalization may be of utility, in particular, for simulating
single neutrino events in a neutrino telescope. Furthermore, the
method makes no assumptions specific to neutrino transport and can
be extended almost straightforwardly to the problem of transport of
high-energy particles other than neutrinos (e.g. hadrons and muons).

Under certain conditions, neutrinos may transform, changing their
flavor via processes like
$\overline{\nu}_ee^-\rightarrow\overline{\nu}_\ell\ell^-$
or $\nu_\ell e^-\rightarrow\nu_e\ell^-$ ($\ell\neq e$),
owing to production and decay of short-lived hadrons ($D, D_s$,
etc.), or (in the mentioned neutrino fields) through reaction chains
like $\nu_\mu\overline{\nu}_\tau\rightarrow\mu^-\tau^+$,
$\tau^+\rightarrow\overline{\nu}_\tau X$, etc.
As it pointed out recently~\cite{Halzen98}, high-energy tau neutrinos
(and antineutrinos) will effectively regenerate in matter, losing
energy, through the charged-current reaction chain
$\nu_\tau N\rightarrow\tau X$, $\tau\rightarrow\nu_\tau X$.
Such mechanisms must be taken into account in data processing from
many future experiments (detecting $\nu_\tau$ events from astrophysical
neutrino oscillations at energies $\gtrsim 1$~PeV~\cite{Tau}, events
with energy release well beyond the Greisen-Zatsepin-Kuz'min
cutoff~\cite{GZK}, etc.). 

Mathematically, the inclusion of processes changing the neutrino
flavor or neutrino energy loss through creation and decay of
short-lived particles leads to a system of transport equations.
The extension of the method to the general system is not
straightforward and demands additional assumptions specific to the
task. However, the case when these contributions may be treated as
corrections presents no special problem. Since this case is rather
common (the neutrino production in the $\nu$-induced hadronic
cascades is a typical example), we brief the corresponding trivial
generalization in appendix~\ref{app:GF}.

To avoid technical complications, in the main text we shall neglect
the (standard and hypothetical) flavor-changing neutrino
interactions%
\footnote[1]{As well as the effects of possible neutrino flavor
                                        mixing.\label{note:mixing}}
and use the simplest ``standard'' scenario for neutrino propagation
described by a \emph{single} transport equation. We shall consider
sufficiently high energies in order to neglect the thermal velocities
of the scatterers in the target medium and to deal with the
one-dimensional theory.
As an illustration, we shall discuss results obtained with some
specific models for the initial spectra of muon neutrinos and
antineutrinos propagating through a normal cold medium.

\protect\section{Method for the solution
                 of the neutrino transport equation}\label{sec:Z}

Let $F_\nu(E,x)$ be the differential energy spectrum of neutrinos at
a column depth $x$ in the medium defined by
\[
x = \int_0^L\rho(L')dL',
\]
where $\rho(L)$ is the density of the medium at a distance $L$ from
the boundary measured along the neutrino beam path. Then, neglecting
the flavor-changing processes mentioned in the introduction,
%%%% (and thermal velocities of the scatterers)
one can derive the following one-dimensional transport equation
\begin{equation}\label{TE}
 \frac{\partial F_\nu(E,x)}{\partial x} = \frac{1}{\lambda_\nu(E)}
 \left[\int_0^1\Phi_\nu(y,E)F_\nu\left(\frac{E}{1-y},x\right)
                                \frac{\d y}{1-y}-F_\nu(E,x)\right],
\end{equation}
with the boundary condition $F_\nu(E,0) = F_\nu^0(E)$. Here
$\lambda_\nu(E)$ is the neutrino interaction length defined by the
equation
\[
\frac{1}{\lambda_\nu(E)}=\sum_T N_T\sigma_{\nu T}^{\mathrm{tot}}(E),
\]
where $N_T$ is the number of scatterers $T$ in 1 g of the medium,
$\sigma_{\nu T}^{\mathrm{tot}}(E)$ is the total cross section for
the $\nu T$ interactions and the sum is over all scatterer types
($T = N,e,\ldots$). The ``regeneration function'' $\Phi_\nu(y,E)$
is defined by
\[
\sum_T N_T\frac{\d\sigma_{\nu T\rightarrow\nu X}(y,E_y)}{\d y} =
\Phi_\nu(y,E)\sum_T N_T\sigma_{\nu T}^{\mathrm{tot}}(E),
\]
where $\d\sigma_{\nu T\rightarrow\nu X}(y,E)/\d y$ is the
differential cross section for the inclusive reaction
$\nu T\rightarrow\nu X$ (with $E$ the initial neutrino energy and
$y$ the fraction of energy lost) and $E_y \equiv E/(1-y)$.

Let us define the effective absorption length $\Lambda_\nu(E,x)$ by
\begin{equation}\label{Lambda}
F_\nu(E,x) = F_\nu^0(E)\exp\left[-\frac{x}{\Lambda_\nu(E,x)}\right].
\end{equation}
As is evident from eq.~(\ref{TE}), $\Lambda_\nu(E,x)>\lambda_\nu(E)$
for any finite $E$ and $x$. Therefore
\begin{equation}\label{Z}
\Lambda_\nu(E,x) = \frac{\lambda_\nu(E)}{1-Z_\nu(E,x)},
\end{equation}
where $Z_\nu(E,x)$ is a positive function (we will call it $Z$ factor
in analogy with the hadronic cascade theory) which contains the
complete information about the neutrino kinetics in matter.

Substituting eqs.~(\ref{Lambda}) and (\ref{Z}) into eq.~(\ref{TE})
and integrating by parts, it is easy to derive the integral equation
for the $Z$ factor:
\begin{equation}\label{ZE}
Z_\nu(E,x) = \frac{1}{x}\int_0^x\int_0^1\eta_\nu(y,E)\Phi_\nu(y,E)
                    \exp\left[-x'D_\nu(E,E_y,x')\right]\d x'\d y,
\end{equation}
with
\[
D_\nu(E,E_y,x) = \frac{1-Z_\nu(E_y,x)}{\lambda_\nu(E_y)}-
                 \frac{1-Z_\nu(E  ,x)}{\lambda_\nu(E  )}
\]
and
\[
\eta_\nu(y,E) = \frac{F_\nu^0(E_y)}{F_\nu^0(E)(1-y)}.
\]
We dwell on eq.~(\ref{ZE}). Although nonlinear, it proves to be
more amenable for an iteration solution than eq.~(\ref{TE}),
considering the smallness of the $Z$ factor and (what is more
important) a model-independent feature of the regeneration
function $\Phi_\nu(y,E)$, namely, its sharp maximum at $y = 0$.

In this section, we assume that the initial spectrum $F_\nu^0(E)$
is a continuous function decreasing at high energies fast enough
so that $0\leq\eta_\nu(y,E)<\infty$ for any $E$ and $0\leq y \leq 1$.
Actually, the neutrino spectra of interest for high-energy neutrino
astrophysics decrease everywhere so fast that $0\leq\eta_\nu(y,E)<1$
for any $E$ and $y > 0$.

We will first look at the case of thin absorbers. One can readily
see that
\[
Z_\nu(E,0)  =  \int_0^1\eta_\nu(y,E)\Phi_\nu(y,E)\d y
         \equiv Z_\nu^0(E).
\]
The approximation $Z_\nu(E,x) = Z_\nu^0(E)$ is usually utilized
when studying the muon neutrino propagation through matter (see e.g.
ref.~\cite{Berezinsky86,Berezinsky90} and references therein).
However, at high energies this approximation becomes too rough even
for "shallow" (as compared to $\lambda_\nu$) depths. Indeed, taking
into account the $\mathcal{O}(x/\lambda_\nu)$ correction yields
\[
Z_\nu(E,x) \approx Z_\nu^0(E)
                        -\frac{x\Delta_\nu^1(E)}{2\lambda_\nu(E)},
\]
where
\begin{eqnarray*}
\Delta_\nu^1(E) & = &-\lambda_\nu(E)\left[\frac{\partial Z_\nu(E,x)}
                     {\partial x}\right]_{x=0} \\
                & = & \int_0^1\eta_\nu(y,E)\Phi_\nu(y,E)\left\{
                      \left[1-Z_\nu^0(E_y)\right]
                      \frac{\lambda_\nu(E)}{\lambda_\nu(E_y)}-
                      \left[1-Z_\nu^0(E  )\right]\right\}\d y.
\end{eqnarray*}
Thus, the approximation $Z_\nu \approx Z_\nu^0$ can only be valid for
\[
\frac{x}{\lambda_\nu(E)} \ll
                  \frac{2Z_\nu^0(E)}{\left|\Delta_\nu^1(E)\right|},
\]
In the general case, the function $\Delta_\nu^1(E)$ is not small in
comparison with $Z_\nu^0(E)$. This can be demonstrated with the
simple model adopted in ref.~\cite{Nicolaidis96}. The authors of
ref.~\cite{Nicolaidis96} assumed that $\Phi_\nu = \Phi_\nu(y)$ is an
energy independent function, $\lambda_\nu(E)\propto E^{-\beta}$ and
$F_\nu^0(E)\propto E^{-(\gamma+1)}$ with energy independent positive
$\beta$ and $\gamma$. All these assumptions are far from reality but
may have a physical sense at super-high energies. For example, owing
to the $\nu N$ interactions, $\beta$ is a monotonically decreasing
function of $E$ changing from about 1 at $E \lesssim 1$~TeV to about
0.4 at $E \gtrsim 1$~PeV~\cite{Gandhi96}; the function $\Phi_\nu(y,E)$
strongly varies with $E$ at all energies, but for $E \gtrsim 1$~PeV
it may be roughly approximated by a scaling function
(see fig.~\ref{f:CS} in sect.~\ref{sec:NumRes}).%
\footnote[2]{However, as fig.~\ref{f:CS}.b suggests, the specific
             parametrization used in ref.~\cite{Nicolaidis96},
             $\Phi_\nu(y)\propto (\mathrm{const}+y)^{-1}$, is too
             rough even at super-high energies.\label{note:param}}
In this model, both $Z_\nu^0$ and $\Delta_\nu^1$ are energy
independent:
\begin{eqnarray*}
Z_\nu^0      & = & \int_0^1(1-y)^\gamma\Phi_\nu(y)\d y, \\
\Delta_\nu^1 & = & \left(1-Z_\nu^0\right)\int_0^1(1-y)^\gamma
                   \left[(1-y)^{-\beta}-1\right]\Phi_\nu(y)\d y.
\end{eqnarray*}
Evidently $\Delta_\nu^1$ can be much larger than $Z_\nu^0$ for a
sufficiently hard initial neutrino spectrum (small $\gamma$)%
\footnote[3]{In the real case, $\Delta_\nu^1(E)$ is nevertheless
             finite because any physical spectrum $F_\nu^0(E)$ has
             a cutoff at some finite energy $E_{\mathrm{cut}}$ and
             therefore $\eta_\nu(y,E) = 0$ at
             $y \geq 1-E/E_{\mathrm{cut}}$.\label{note:cutoff}}.

It is not a hard task to derive the
$\mathcal{O}\left(\left(x/\lambda_\nu\right)^k\right)$ corrections
for $k = 2,3,\ldots$, but as a result we will get an asymptotic
expansion with coefficient functions, $\Delta_\nu^k(E)$, increasing
fast with $k$.  The region of applicability of this expansion proves
to be very limited and decreases fast with increasing energy.

Now, let us consider a way to solve eq.~(\ref{ZE}) for any depth
and energy. We will use an iteration algorithm. Let $n$ label the
iteration number. Then we define
\begin{equation}\label{Dn}
D_\nu^{(n)}(E,E_y,x)=\frac{1-Z_\nu^{(n)}(E_y,x)}{\lambda_\nu(E_y)}-
                     \frac{1-Z_\nu^{(n)}(E  ,x)}{\lambda_\nu(E  )}
\end{equation}
and
\begin{equation}\label{Zn}\hspace{-8mm}
Z_\nu^{(n+1)}(E,x) = \frac{1}{x}\int_0^x\int_0^1\eta_\nu(y,E)
\Phi_\nu(y,E)\exp\left[-x'D_\nu^{(n)}(E,E_y,x')\right]\d x'\d y.
\end{equation}
Due to the mentioned sharp maximum of $\Phi_\nu(y,E)$, the main
contribution into the integral over $y$ on the right side of
eq.~(\ref{Zn}) comes from the lover limit neighborhood. But
$D_\nu(E,E_y,x)\rightarrow 0$ as $y \rightarrow 0$ and thus
the algorithm is robust in respect to choosing the zero
approximation. The simplest choice is $Z_\nu^{(0)}(E,x) = 0$.
Therefore
\begin{equation}\label{D0}
D_\nu^{(0)}(E,E_y,x)  =
\frac{1}{\lambda_\nu(E_y)}-\frac{1}{\lambda_\nu(E)}
                    \equiv \mathcal{D}_\nu(E,E_y),
\end{equation}

The algorithm (\ref{Dn}--\ref{D0}) is formally applicable for
arbitrary decreasing initial spectra. It is however clear that the
softer the initial spectrum, the better the convergence of the
algorithm. In the next section, we show that the algorithm converges
very fast for realistic initial spectra and has no restrictions in
depth or energy. Moreover, even the first approximation,
\begin{equation}\label{Z0}
Z_\nu^{(1)}(E,x) = \int_0^1\eta_\nu(y,E)\Phi_\nu(y,E)
\left[\frac{1-\e^{-x\mathcal{D}_\nu(E,E_y)}}
                  {x\mathcal{D}_\nu(E,E_y)}\right]\d y,
\end{equation}
proves to be quite accurate. It has the correct asymptotic behavior
both in energy and depth and can thus be used for an analytical or
numerical evaluation of the $Z$ factor with a not-too-big error.

Assuming that $\lambda_\nu(E)$ is a decreasing function%
\footnote[4]{For a normal medium, this is true for all neutrino
             flavors except $\overline{\nu}_e$
             (see ref.~\cite{Gandhi96}).\label{note:exception}}
and therefore $\mathcal{D}_\nu(E,E_y) > 0$ for $y > 0$,
one can prove that
\begin{itemize}
\item $Z_\nu^{(1)}(E,x) \leq Z_\nu^0(E)$ at any $E$ and $x \geq 0$,
      and
\item $Z_\nu^{(1)}(E,x)\rightarrow 0$ as $x\rightarrow\infty$ at
      any $E$.
%%% $|\partial Z_\nu^{(1)}(E,x)/\partial x|$ increases with energy.
\end{itemize}
The latter signifies that neutrino ``regeneration'' due to the
inclusive reactions $\nu T\rightarrow\nu X$ becomes negligible at
sufficiently large depths. This conclusion remains true for the exact
solution of eq.~(\ref{ZE}), under rather general assumptions about
the behavior of the initial neutrino spectrum and cross sections.

\protect\section{Numerical Illustration and Discussion}
\label{sec:NumRes}

Below, we will consider only muon neutrinos and antineutrinos
propagating through a normal medium. But, with obvious reservations,
the results that follow also hold for electron neutrinos.
In this calculation, we shall neglect neutrino scattering off electrons
(thus we exclude electron antineutrinos from our consideration) as well
as the neutrino ``recreation'' in the $\nu$-induced cascades.
For further simplification, we shall deal with an isoscalar medium
neglecting nuclear effects.

To calculate the differential $\nu_\mu N$ and $\overline{\nu}_\mu N$
cross sections we use the approach of ref.~\cite{Gandhi96} based on
the renormalization-group-improved parton model and new experimental
information about the quark structure of the nucleon (see
appendix~\ref{app:NNCS}). Various versions of different sets of parton
density functions $q(\hat{x},Q^2)$ are now collected in a large CERN
program library PDFLIB~\cite{PDFLIB97}; they can be simply accessed
by setting few parameters to choose the desired version. In this
calculations, we selected, following ref.~\cite{Gandhi96}, the third
version of the CTEQ collaboration model~\cite{Lai95}, because it is
characterized by a particularly suitable extrapolation at very low
Bjorken $\hat{x}$. The evolution in $Q^2$ is realized by
next-to-leading order Altarelli-Parisi equations from initial
$Q_0^2 = 2.56$~GeV${}^2$.

\begin{figure}[h]
\centering\mbox{\epsfig{file=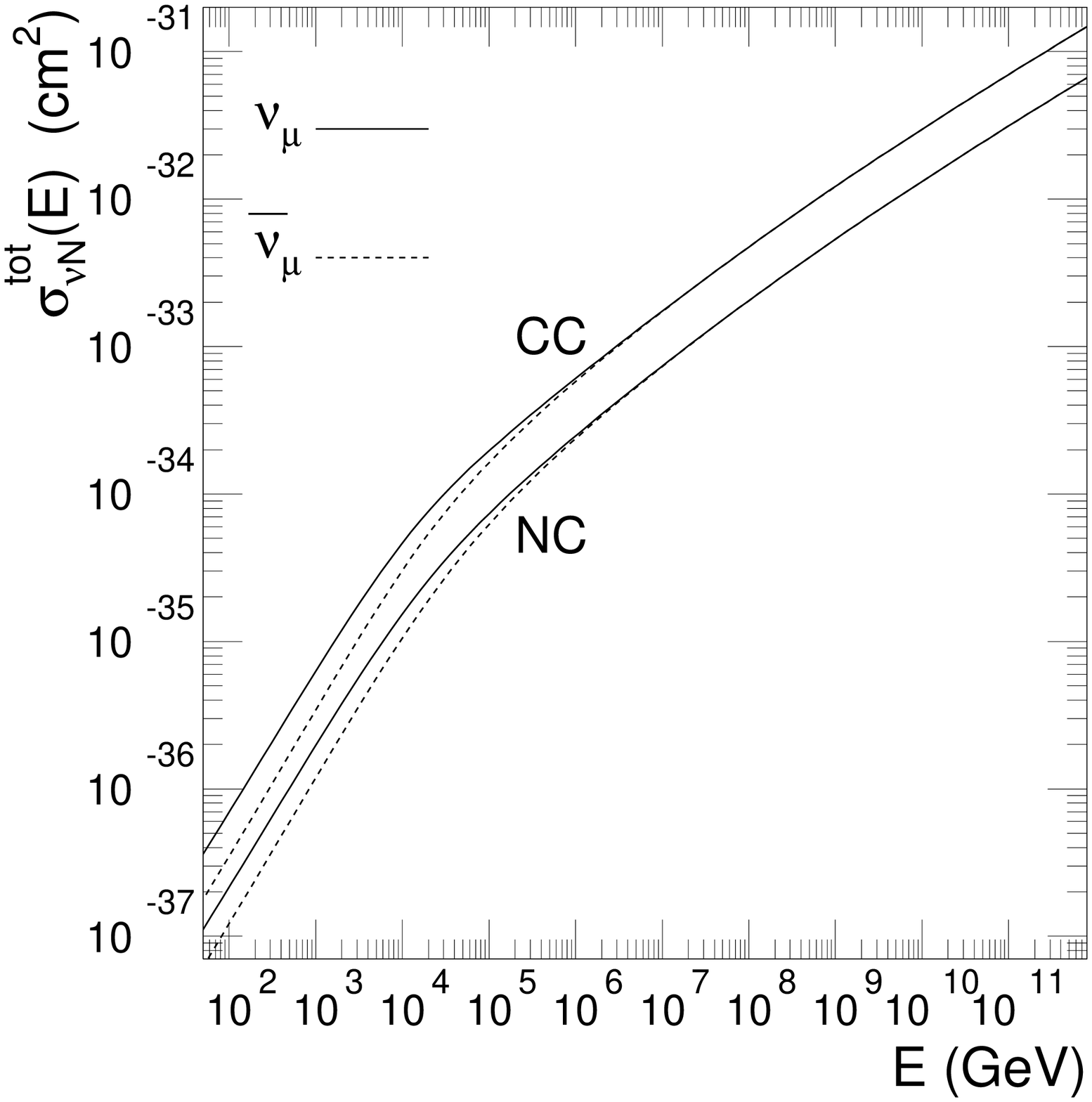,width=6.8cm,height=6.8cm}
                \epsfig{file=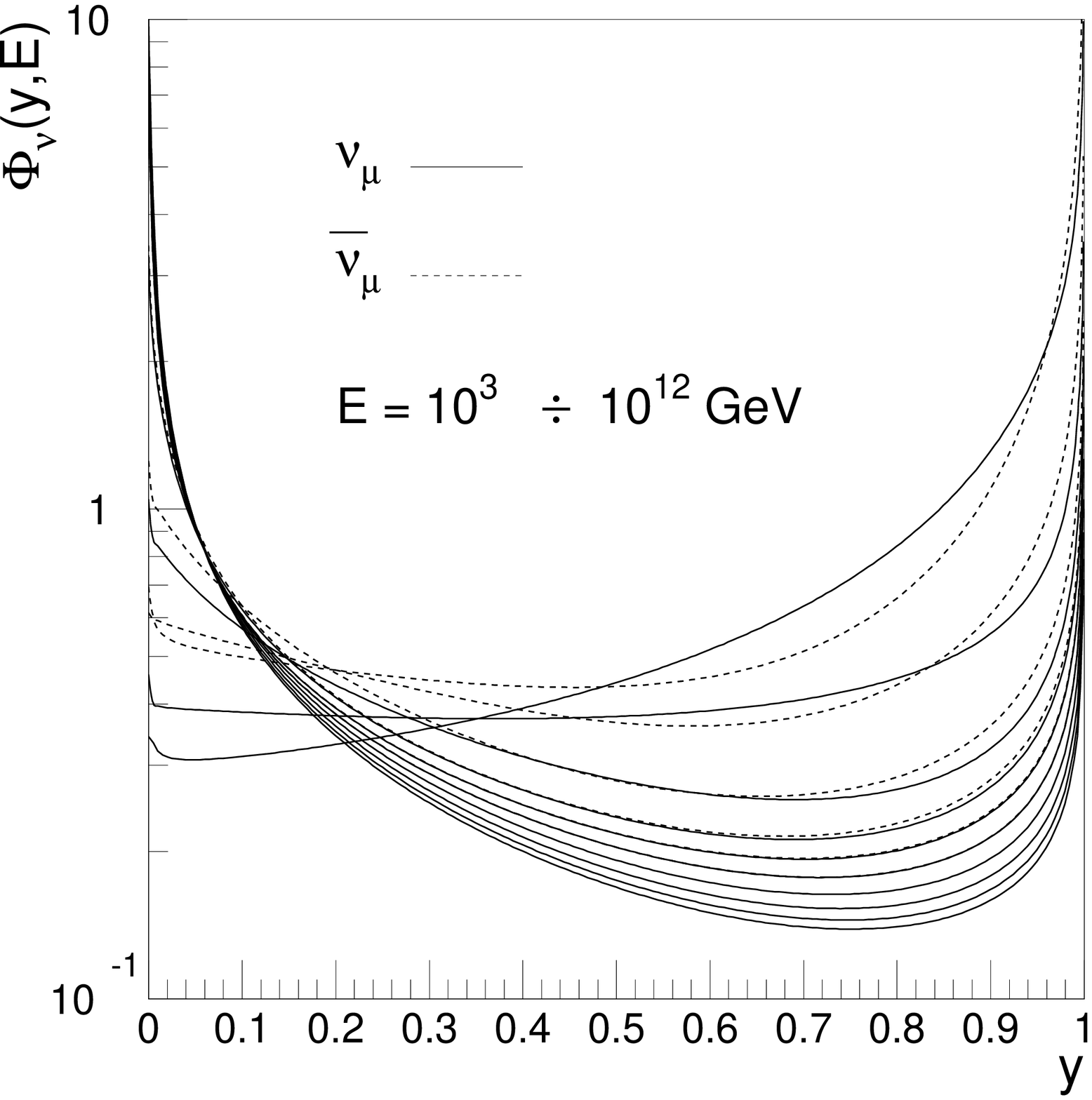,width=6.8cm,height=6.8cm}}
\protect\caption{Total (CC and NC) $\nu_\mu N$ and
                 $\overline{\nu}_\mu N$ cross sections vs energy (a)
                 and regeneration functions $\Phi_{\nu_\mu}(y,E)$ and
                 $\Phi_{\overline{\nu}_\mu}(y,E)$ vs $y$ for
                 $E = 10^k$~GeV [$k = 3, 4, \ldots, 12$ from top to
                 bottom] (b).
\label{f:CS}}
\end{figure}

The total cross sections for CC and NC inelastic scattering of muon
neutrinos and antineutrinos off an isoscalar nucleon are shown in
fig.~\ref{f:CS}.a as the solid ($\nu_\mu$) and dashed
($\overline{\nu}_\mu$) curves.  Fig.~\ref{f:CS}.b shows the
regeneration functions $\Phi_{\nu_\mu}(y,E)$ (solid curves) and
$\Phi_{\overline{\nu}_\mu}(y,E)$ (dashed curves) versus $y$ for
several values of $E$ ($10^3$ to $10^{12}$~GeV).

%--------------------------------------------------------------------
%--------------------------------------------------------------------

At all energies, our calculation for the cross sections agrees with
the result of ref.~\cite{Gandhi96} within a few percent accuracy; the
insignificant difference near the resonance region is due mainly to
differences in the adopted values for the electroweak parameters
($W/Z$ boson masses, $t$ quark mass, Weinberg angle, etc.)%
\footnote[5]{In our calculation, all these parameters were updated
             according to the PDG data~\cite{PDG96}.}
and, at superhigh energies, to the top sea contribution neglected in
ref.~\cite{Gandhi96}. As one can see from the figures, the $\nu_\mu$
and $\overline{\nu}_\mu$ scatterings become indistinguishable for
$E \gtrsim 1$~PeV.

We use the following model for the initial neutrino spectrum:
\begin{equation}\label{F0}
F_\nu^0(E) = K\left(\frac{E_0}{E}\right)^{\gamma+1}
                       \left(1+\frac{E}{E_0}\right)^{-\alpha}
                       \phi\left(\frac{E}{E_{\mathrm{cut}}}\right),
\end{equation}
where $K$, $\gamma$, $\alpha$, $E_0$ and $E_{\mathrm{cut}}$ are
parameters and $\phi(t)$ is a function equal to 0 at $t\geq1$ and 1
at $t \ll 1$. Varying the parameters in eq.~(\ref{F0}), we can
approximate many models for the neutrino fluxes expected from the
known astrophysical sources.  Technically, the function $\phi(t)$
serves to avoid an extrapolation of the cross sections to the
ultrahigh energy region for which our knowledge of the parton density
functions becomes doubtful. For realistic values of the parameters
$\gamma$, $\alpha$ and $E_0$, the explicit form of $\phi(t)$ is
of no importance if one is interested in the energy range
$E \ll E_{\mathrm{cut}}$. In fact, $\phi(t)$ may be treated as the
real physical cutoff of the spectrum determined by the energetics of
the neutrino source or by neutrino interactions with the cosmic
backgrounds. In the present calculations, we adopt (without serious
physics arguments)
$\phi(t)=1/\left[1+\tan\left(\pi t/2\right)\right]$ ($t < 1$) and
$E_{\mathrm{cut}} = 3\times10^{10}$~GeV.

%--------------------------------------------------------------------
\begin{figure}[thb]
\centering\mbox{\epsfig{file=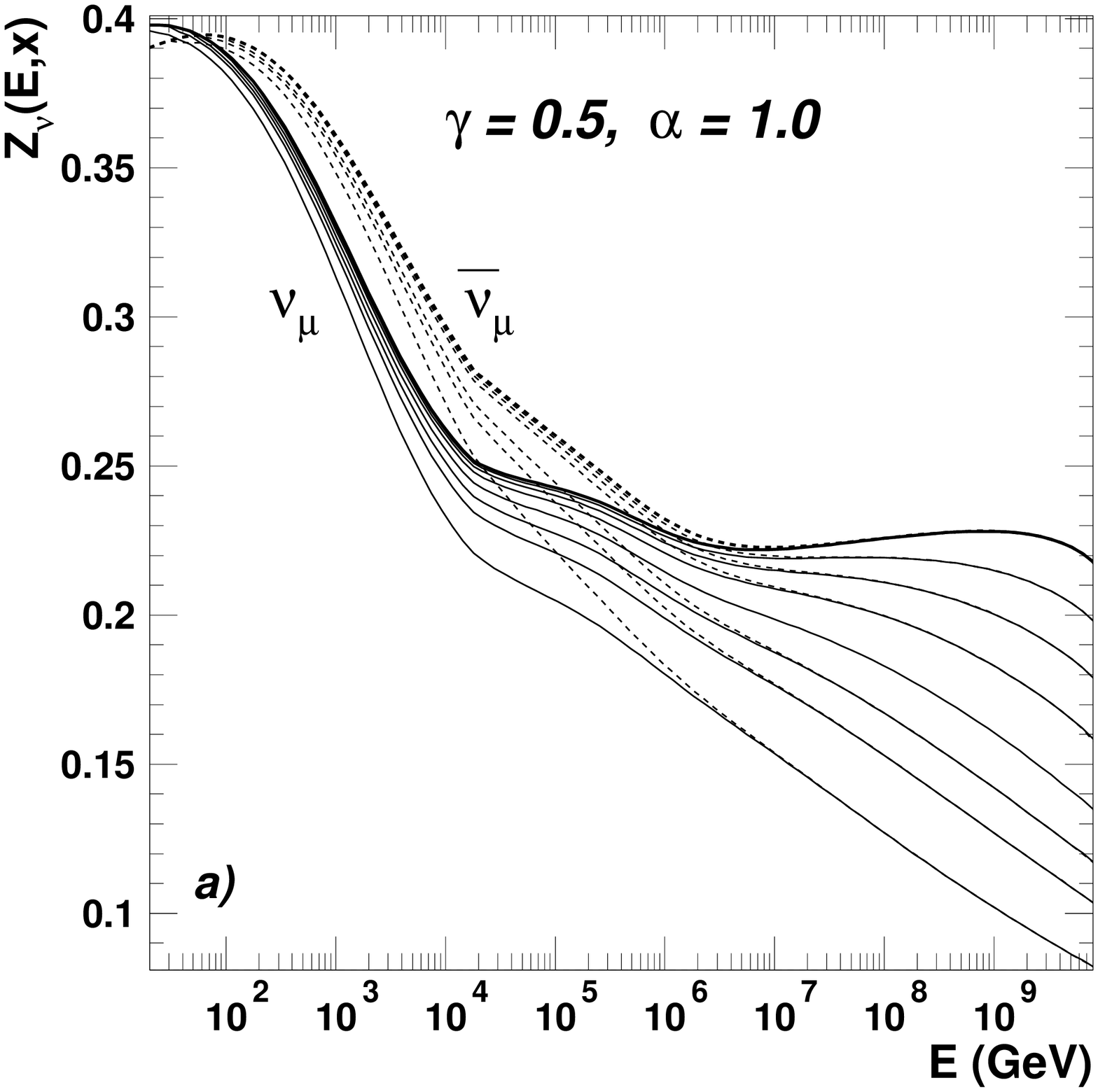,width=6.8cm,height=7.0cm}
                \epsfig{file=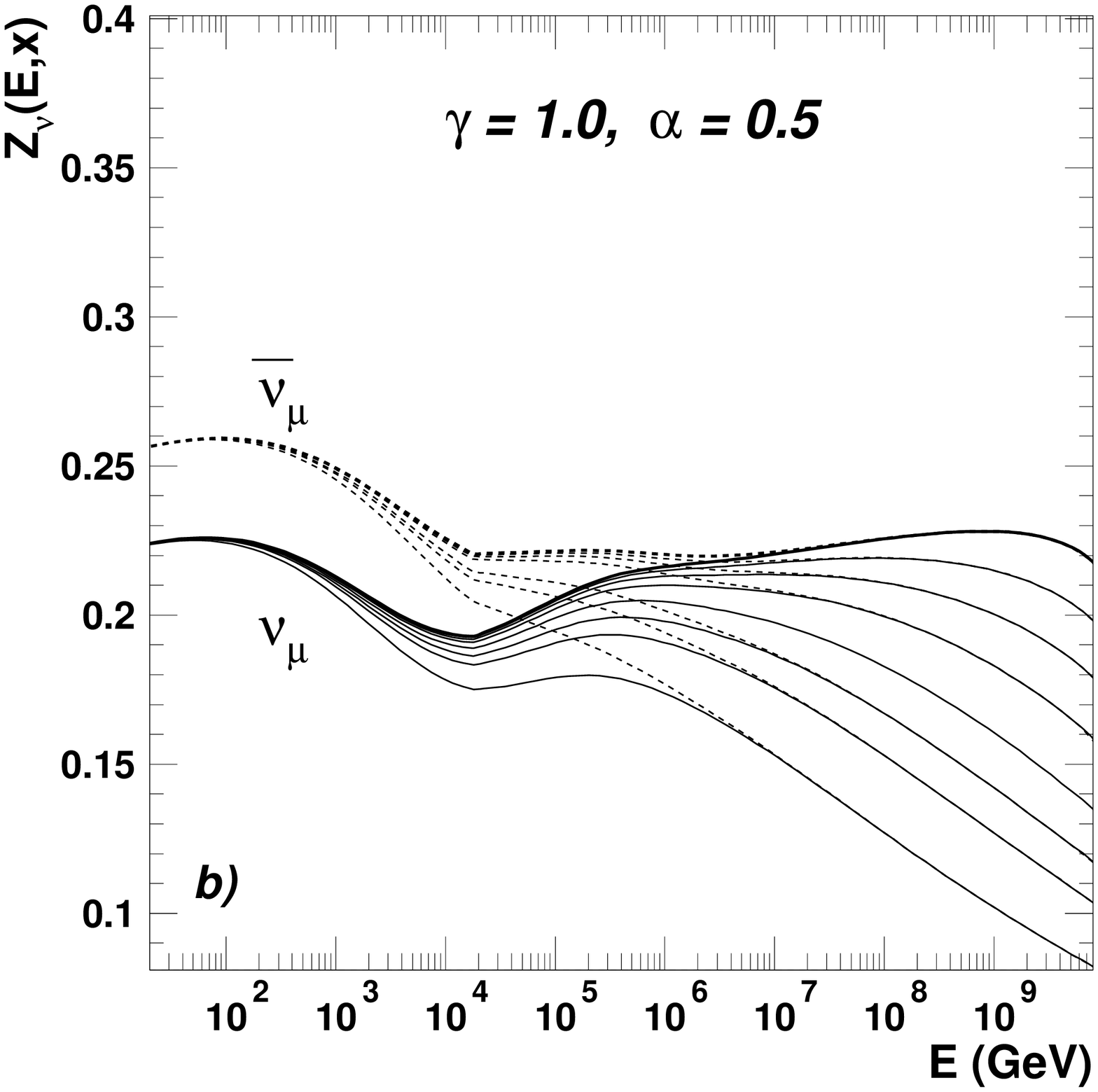,width=6.8cm,height=7.0cm}}
\centering\mbox{\epsfig{file=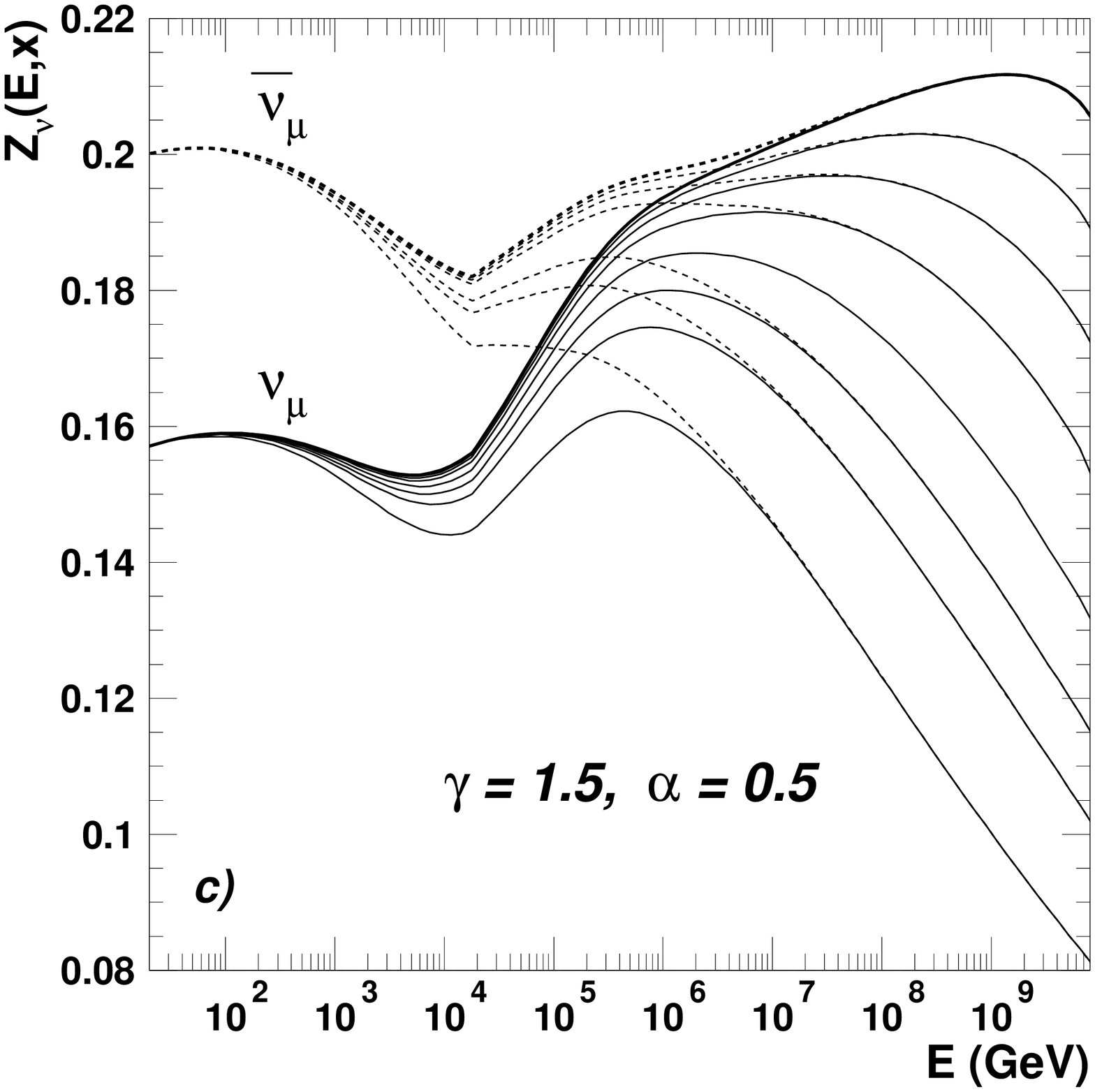,width=6.8cm,height=7.0cm}
                \epsfig{file=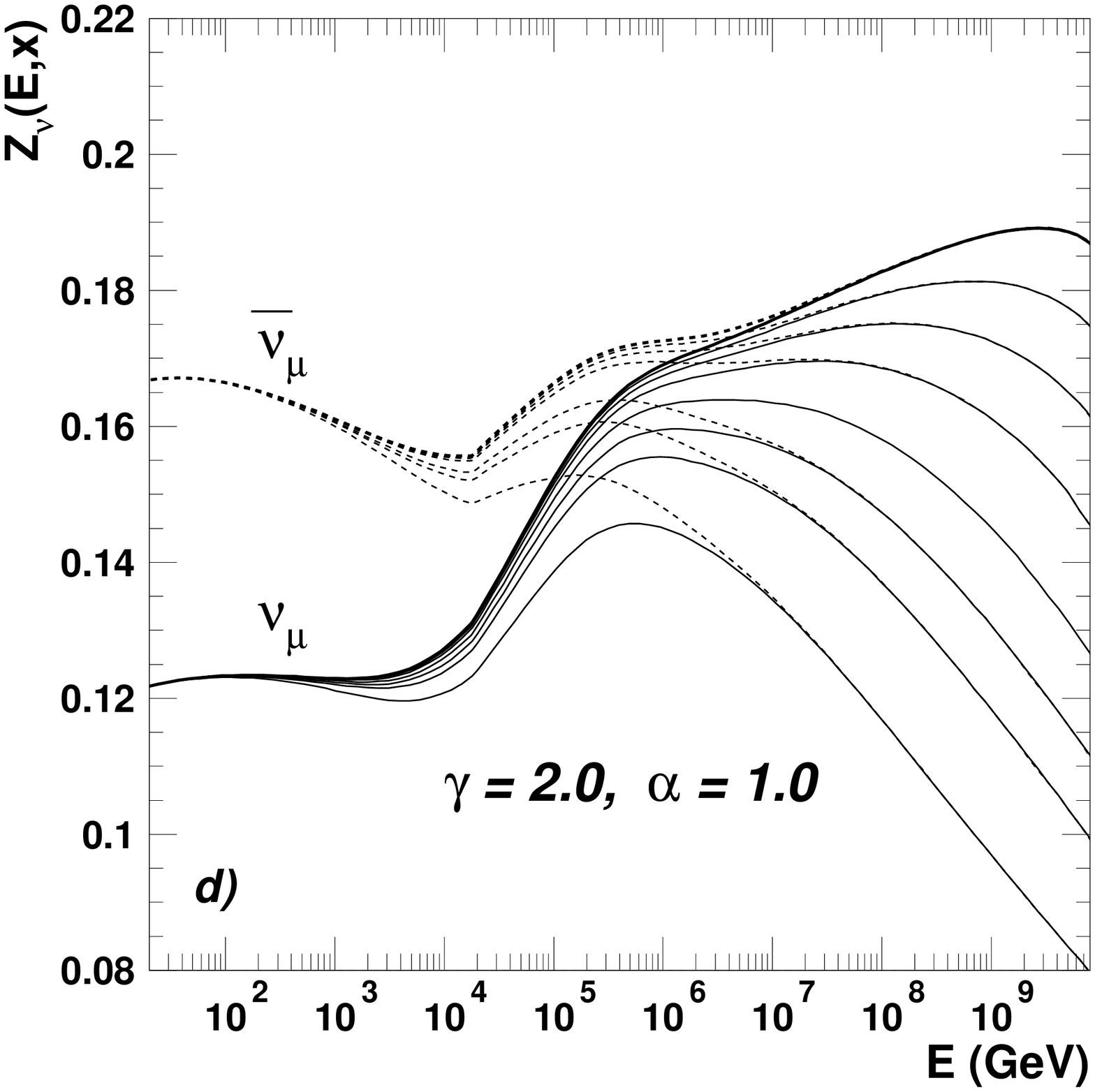,width=6.8cm,height=7.0cm}}
\protect\caption{$Z$ factors, $Z_{\overline{\nu}_\mu}(E,x)$ and
                 $Z_{\nu_\mu}(E,x)$ vs energy for the initial spectra
                 (\protect\ref{F0}), calculated with four different
                 sets of $\gamma$ and $\alpha$ and with $E_0 = 1$~PeV
                 for depths $x=x_\oplus/k$ [$k = 1,2,3,5,10,20,50$
                 from bottom to top] and $x=0$
                 (the largest $Z$ factors).
\label{f:ZE}}
\end{figure}
%--------------------------------------------------------------------

Fig.~\ref{f:ZE} shows the energy dependence of the $Z$ factors,
$Z_{\nu_\mu}(E,x)$ (solid curves) and $Z_{\overline{\nu}_\mu}(E,x)$
(dashed curves) for various depths, from $x=0$ to $x=x_\oplus$
(where $x_\oplus \approx 1.1\times10^{10}$~g/cm${}^2$ is the depth of
the earth along the diameter), for the initial spectra
(\protect\ref{F0}) calculated with $\gamma = 0.5$, $\alpha = 1$ (a),
$\gamma = 1$, $\alpha = 0.5$ (b), $\gamma = 1.5$, $\alpha = 0.5$ (c)
and $\gamma = 2$, $\alpha = 1$ (d).  In all cases we used $E_0 =
1$~PeV.  The calculations were made in the fourth order of the
iteration procedure described in sect.~\ref{sec:Z}. For all the
spectra under discussion, for 10~GeV $\leq E \leq 10^{10}$~GeV and $0
\leq x \leq x_\oplus$, the maximum difference between
$Z_\nu^{(1)}(E,x)$ and $Z_\nu^{(2)}(E,x)$ is about 4\%; the value
$\left|Z_\nu^{(3)}/Z_\nu^{(2)}-1\right|$ is less than
$2\times10^{-3}$ and $\left|Z_\nu^{(4)}/Z_\nu^{(3)}-1\right|$ is less
than the precision of the numerical integration and interpolation
(about $10^{-5}$) adopted in our calculations. After the tests with
many models for the initial spectrum, we conclude that the
convergence of the algorithm is very good and that even the first
approximation, $Z_\nu^{(1)}(E,x)$, has an accuracy quite sufficient
for the majority of applications of the theory.

As it is clear from fig.~\ref{f:ZE}, the shape of the $Z$ factors is
very dependent from the initial spectrum. This is a positive fact for
neutrino astronomy, since it gives, at least in principle, the
possibility to reconstruct the initial neutrino spectrum from the
measured energy spectrum and angular distribution of the neutrino
induced muon events in a neutrino telescope.

At comparatively low energies (except for unrealistically hard
spectra like the one used in fig.~\ref{f:ZE}.a), the $Z$ factors for
antineutrinos exceed the ones for neutrinos. Considering the inequality
$\lambda_{\overline{\nu}_\mu}(E)>\lambda_{\nu_\mu}(E)$, one can
conclude that 
\[
\Lambda_{\overline{\nu}_\mu}(E,x) > \Lambda_{\nu_\mu}(E,x)
\]
for any depth. In the multi-PeV energy region and above, the $Z$ factors
(and effective absorbtion lengths) are identical for $\nu_\mu$ and
$\overline{\nu}_\mu$. The difference between the shapes of
$Z_{\nu_\mu}(E,x)$ and $Z_{\overline{\nu}_\mu}(E,x)$ is almost
depth-independent and becomes more important for steep initial spectra.
This behavior may be understood from an analysis of the shapes of the
total cross sections and regeneration functions for $\nu_\mu$ and
$\overline{\nu}_\mu$ (fig.~\ref{f:CS}).

At any fixed energy, the $Z$ factors monotonically decrease with
increasing depth and the inequality $Z_\nu(E,x) < Z_\nu^0(E)$ takes
place for any $x > 0$. This effect leads to significant decrease of
the neutrino event rates in comparison with those estimated in the
``standard'' approximation $Z_\nu \approx Z_\nu^0$; the latter works
only at low energies, when the shadow effect is by itself small (that
is when the medium is almost transparent for neutrinos). Although
these conclusions were derived from particular models for the initial
neutrino spectrum, cross sections, and medium, actually they are
highly general and model-independent.
%%%%%%%%%%%%%%%%%%%%%%%%%%%%%%%%%%%%%%%%%%%%%%%%%%%%%%%%%%%%%%%
%% Moreover, similar effects take place in many problems of
%% high-energy particle transport.
%%%%%%%%%%%%%%%%%%%%%%%%%%%%%%%%%%%%%%%%%%%%%%%%%%%%%%%%%%%%%%%

\begin{figure}[htb]
\centering\mbox{\epsfig{file=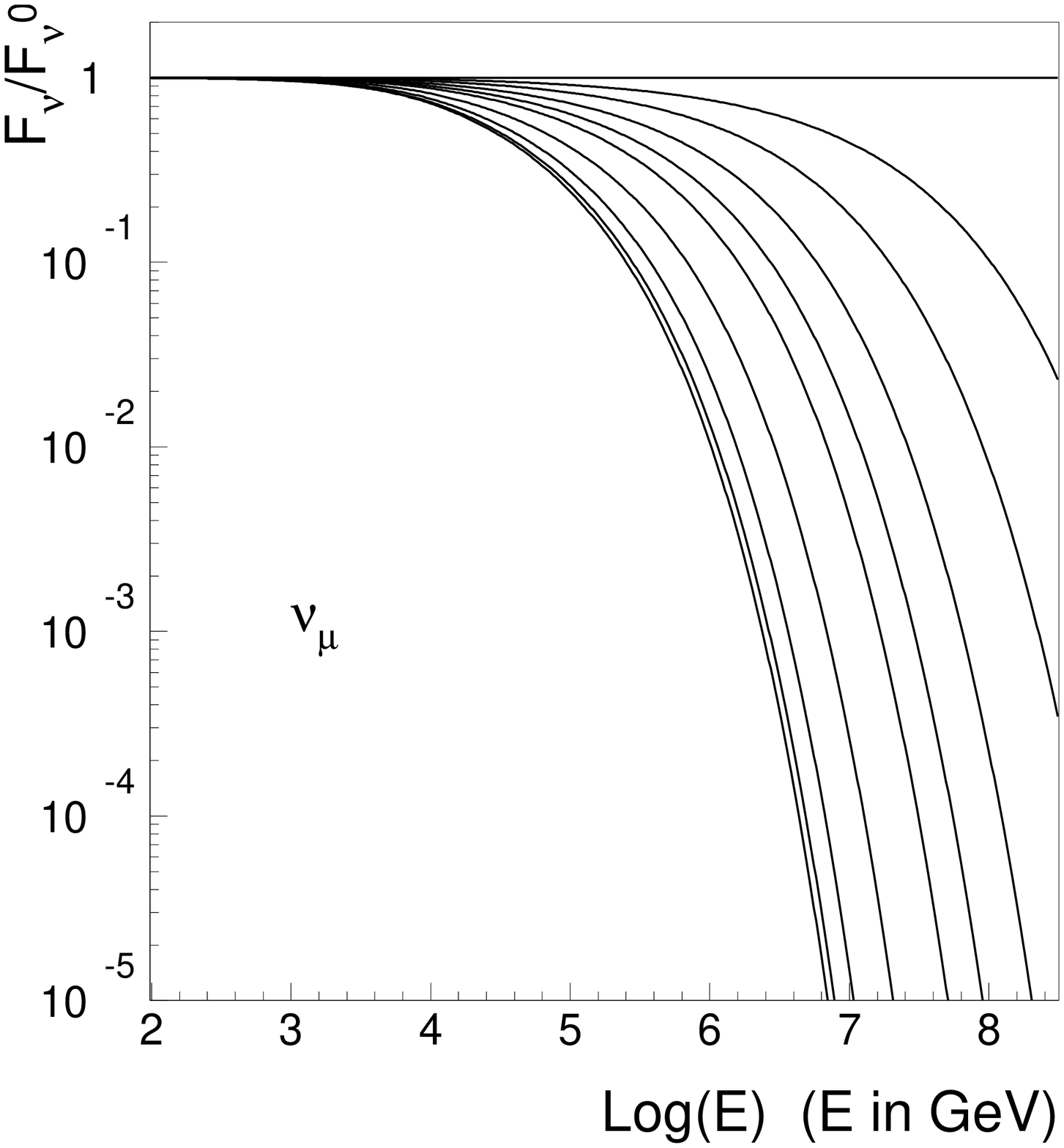,width=6.8cm,height=6.8cm}
                \epsfig{file=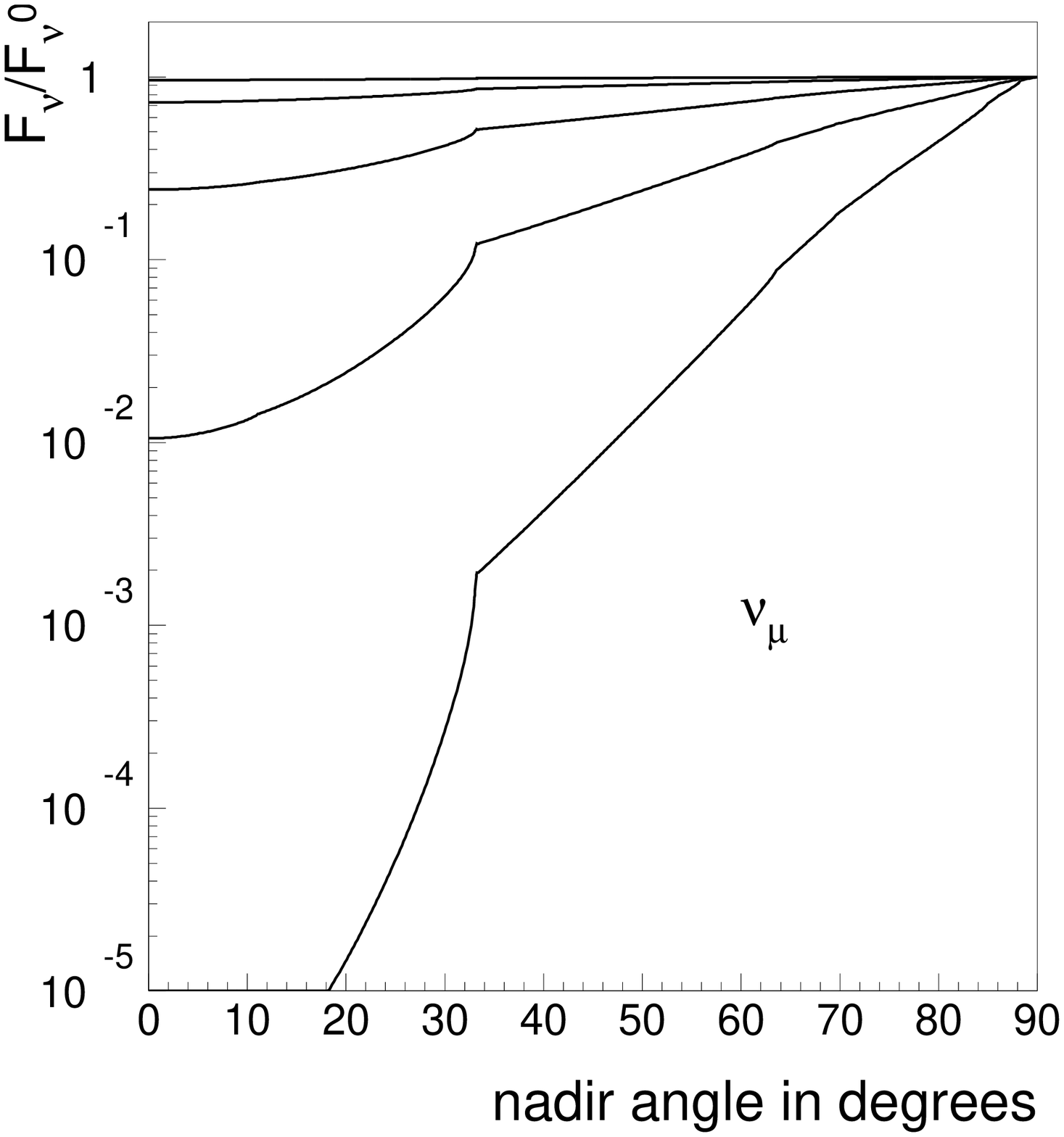,width=6.8cm,height=6.8cm}}
\protect\caption{Neutrino penetration coefficient in the earth for
                 the quasi-power-law initial spectrum with
                 $\gamma=0.7$ as a function of $E$ at fixed
                 $\vartheta$ [$0^\circ$ to $90^\circ$ from bottom to
                 top with steps of $10^\circ$] (a) and as a function
                 of $\vartheta$ for $E=10^k$~GeV [$k=3,4,\ldots,7$
                 from top to bottom] (b).
\label{f:Shad}}
\end{figure}

In fig.~\ref{f:Shad} we present the penetration coefficient,
$\exp\left[-x/\Lambda_\nu(E,x)\right]$, in the earth for muon
neutrinos with initial spectrum (\ref{F0}) calculated with
$\gamma = 0.7$ and $\alpha = 0$ (``quasi-power-law'' spectrum).  The
results are presented as a function of $E$ for several nadir angles
($\vartheta$) in fig.~\ref{f:Shad}.a and as a function of $\vartheta$
for several values of $E$ in fig.~\ref{f:Shad}.b.

%--------------------------------------------------------------------
%--------------------------------------------------------------------
 
To evaluate the depth, $x$, as a function of $\vartheta$, we used the
density profile of the earth, $\rho(L)$, given in ref.~\cite{Gandhi96}.
The kinks in fig.~\ref{f:Shad}.b are due to the layered structure of
the earth.

\ack %\acknowledgements

The work of V.\,N. supported in part by the Ministry of General and
Professional Education of Russian Federation, within the framework of
the Program ``Universities of Russia -- Basic Researches'',
Grant No.~728.
We want to acknowledge helpful and constructive discussions with
Bianca Monteleoni whose encouragement stimulated this work. We also
thank Ina Sarcevic for providing us with the Fortran code which was
used to test our calculations of the $\nu N$ cross sections.
V.\,N. acknowledges the hospitality of INFN, Sezione di Firenze,
where this research was completed.

\clearpage

\appendix

\protect\section{Monochromatic initial spectrum}
\label{app:MIS}

Let us consider an initial ``spectrum'' of the form $\delta(E-E_0)$
with a fixed parameter $E_0$. In this appendix, we will show how this
monochromatic spectrum transforms at a depth $x$ in a medium.
Let us denote the transformed spectrum by $G_\nu(E_0;E,x)$. This
function must satisfy eq.~(\ref{TE}) and simple considerations
suggest the following ansatz:
\begin{equation}\label{ansatz}
G_\nu(E_0;E,x) = \left[\delta(E-E_0)+\frac{\theta(E_0-E)}{E}
                    \psi_\nu(E_0;E,x)\right]\e^{-x/\lambda_\nu(E_0)},
\end{equation}
where the term with $\delta$ function describes absorption of the
initial (``parent'') neutrinos of energy $E_0$ and the next term
-- the creation and propagation of secondary neutrinos with energy
$E < E_0$. Substituting eq.~(\ref{ansatz}) into eq.~(\ref{TE})
yields
\begin{eqnarray}\label{psiEq}
\frac{\partial\psi_\nu(E_0;E,x)}{\partial x}
& = &  \frac{1}{\lambda_\nu(E)}\left[\int_0^{y_0}\Phi_\nu(y,E)
       \psi_\nu\left(E_0;E_y,x\right)\d y
      +\Omega_\nu\left(E,E_0\right)\right] \nonumber\\
&   & +\mathcal{D}_\nu(E,E_0)\psi_\nu\left(E_0;E,x\right),
\qquad \psi_\nu\left(E_0;E,0\right) = 0,
\end{eqnarray}
where $\Omega_\nu\left(E,E_0\right) =
\left(1-y_0\right)\Phi_\nu\left(y_0,E\right)$,
$y_0 \equiv 1-E/E_0 < 1$, and $\mathcal{D}_\nu(E,E_0)$ is
defined by eq.~(\ref{D0})

Let us seek the solution to eq.~(\ref{psiEq}) in the form
\begin{equation}\label{psi}
\psi_\nu(E_0;E,x) = \Omega_\nu\left(E,E_0\right)
                    \int_0^x\exp\left[\,\int_{x'}^x\frac{\d x''}
   {\mathcal{L}_\nu(E_0;E,x'')}\right]\frac{\d x'}{\lambda_\nu(E)},
\end{equation}
\begin{equation}\label{L}
\frac{1}{\mathcal{L}_\nu(E_0;E,x)} = \frac{1}{\lambda_\nu(E_0)}-
                 \frac{1-\mathcal{Z}_\nu(E_0;E,x)}{\lambda_\nu(E)},
\end{equation}
with $\mathcal{Z}_\nu(E_0;E,x)$ an unknown positive function. After
direct substitution of eqs.~(\ref{psi}) and (\ref{L}) into
eq.~(\ref{psiEq}) we have
\begin{equation}\label{calZ}
\mathcal{Z}_\nu(E_0;E,x)\psi_\nu(E_0;E,x) =
       \int_0^{y_0}\Phi_\nu(y,E)\psi_\nu\left(E_0;E_y,x\right)\d y.
\end{equation}
Clearly, $\mathcal{Z}_\nu(E_0;E,x) \rightarrow 0$ and
$\psi_\nu(E_0;E,x) \rightarrow x\Phi_\nu(0,E_0)/\lambda_\nu(E_0)$ as
$E \rightarrow E_0$ for any $x$.

The new ``$Z$ factor'', $\mathcal{Z}_\nu(E_0;E,x)$, can be found from
eqs.~(\ref{psi}--\ref{calZ}) by an iteration algorithm similar
to the algorithm described in sect.~\ref{sec:Z}.
Putting $\mathcal{Z}_\nu = 0$ as a zero approximation we have
\begin{equation}\label{psi0}
\psi_\nu^{(0)}(E_0;E,x) = \frac{\Omega_\nu\left(E,E_0\right)}
                              {\lambda_\nu(E)\mathcal{D}_\nu(E,E_0)}
\left[\e^{x\mathcal{D}_\nu(E,E_0)}-1\right]
\end{equation}
and subsequently
\begin{equation}\label{calZ1}
\mathcal{Z}_\nu^{(1)}(E_0;E,x) = \int_0^{y_0}\Phi_\nu(y,E)
\left[\frac{\psi_\nu^{(0)}\left(E_0;E_y,x\right)}
           {\psi_\nu^{(0)}\left(E_0;E  ,x\right)}\right]\d y.
\end{equation}
The next steps of the algorithm are quite obvious so there is no need
to write out the corresponding cumbersome formulas here.

Let us briefly sketch the leading approximations for $\psi_\nu$ and
$\mathcal{Z}_\nu$, since they contain the main features of the exact
solution. As it is seen from eq.~(\ref{psi0}), for every $E < E_0$
there is a depth
\[
x_*(E_0,E) = \frac{1}{\mathcal{D}_\nu(E,E_0)}
            \ln\frac{\lambda_\nu(E)}{\lambda_\nu(E_0)}
\]
at which the flux of neutrinos of energy $E$ reaches the maximum.
The function $x_*(E_0,E)$ increases when $E$ decreases and tends to
the minimum, $\lambda_\nu(E_0)$, as $E \rightarrow E_0$. At any
finite depth, secondary neutrinos ``remember'' about their parents
(the $E_0$ dependence does not disappear with increasing depth). Due
to the nontrivial shape of the regeneration function $\Phi_\nu$ (see
fig.~\ref{f:CS}.b), the spectrum of secondary neutrinos is rather
complex and transforms fast with depth.

For $x\ll\lambda(E_0)$, the function $\psi_\nu^{(0)}$ behaves as
$x\Omega_\nu\left(E,E_0\right)/\lambda_\nu(E)$. Therefore
\[
\mathcal{Z}_\nu^{(1)}(E_0;E,0) = \int_0^{y_0}\Phi_\nu(y,E)\left[
\frac{\Omega_\nu\left(E_y,E_0\right)\lambda_\nu(E)}
     {\Omega_\nu\left(E  ,E_0\right)\lambda_\nu(E_y)}\right]\d y.
\]
Taking into account that $\lambda_\nu(E) > \lambda_\nu(E_0)$ for
$E < E_0$ (see footnote~\ref{note:exception}), we get the asymptotic
behavior of $\mathcal{Z}_\nu^{(1)}$ for $x\rightarrow\infty$:
\begin{eqnarray*}
\mathcal{Z}_\nu^{(1)}(E_0;E,x) & \sim & \int_0^{y_0}\Phi_\nu(y,E)
     \left[\frac{\Omega_\nu\left(E_y,E_0\right)}
                {\Omega_\nu\left(E  ,E_0\right)}\right]
     \left[\frac{\lambda_\nu(E  )-\lambda_\nu(E_0)}
                {\lambda_\nu(E_y)-\lambda_\nu(E_0)}\right] \\ &   &
\times\exp\left[-x\mathcal{D}_\nu(E,E_y)\right]\d y\rightarrow 0.
\end{eqnarray*}

With the function $\psi_\nu(E_0;E,x)$ in hand, we can obtain the
solution to the transport equation~(\ref{TE}) for \emph{any} initial
spectrum $F_\nu^0(E)$. Indeed, multiplying eq.~(\ref{ansatz}) by
$F_\nu^0(E_0)$ and integrating over $E_0$, we have
\begin{eqnarray}\label{F}
F_\nu(E,x) & = & \int_0^\infty F_\nu^0(E_0)G_\nu(E_0;E,x)\d E_0
                                                         \nonumber \\
           & = & F_\nu^0(E)\e^{-x/\lambda_\nu(E)}
      +\int_E^\infty F_\nu^0(E_0)\psi_\nu(E_0;E,x)
       \e^{-x/\lambda_\nu(E_0)}\frac{\d E_0}{E}.
\end{eqnarray}
The first term on the right side of eq.~(\ref{F}) describes neutrino
absorption and the second the neutrino regeneration due to energy
loss through the reactions $\nu T\rightarrow\nu X$.
Eq.~(\ref{F}) is in fact equivalent to eq.~(\ref{Lambda}) but, when
the function $\psi_\nu(E_0;E,x)$ is known, eq.~(\ref{F}) becomes much
more convenient for calculations because $\psi_\nu$ is independent
from the initial spectrum%
\footnote[6]{However, eq.~(\ref{F}) has one evident technical drawback.
             To use it, one must calculate 3-dimensional arrays which
             are hard to interpolate due to the very strong dependence
             of $\psi_\nu$ and $\mathcal{Z}_\nu$ from their arguments.
             From this point of view, the algorithm described in
             sect.~\ref{sec:Z} is of course simpler.\label{note:a1}}.
Due to the mentioned equivalence, we can get a useful representation
for the $Z$ factor in terms of the function $\psi_\nu$:
\begin{equation}\label{Zpsi}\hspace{-8mm}
Z_\nu(E,x) = \frac{\lambda_\nu(E)}{x}
\ln\left[1+\int_0^1\eta_\nu(y,E)\psi_\nu(E_y;E,x)
                \e^{-x\mathcal{D}_\nu(E,E_y)}\frac{\d y}{1-y}\right].
\end{equation}
It should be noted that the $Z$ factor calculated in the $n$-th
approximation using the algorithm (\ref{Dn}--\ref{Zn}) agrees only
numerically rather than analytically with that calculated from
eq.~(\ref{Zpsi}), using the iteration algorithm for $\psi_\nu$.
In particular, substituting $\psi_\nu = \psi_\nu^{(0)}$ into
eq.~(\ref{Zpsi}) yields
\[
Z_\nu(E,x)  =  \frac{\lambda_\nu(E)}{x}
       \ln\left[1+\frac{xZ_\nu^{(1)}(E,x)}{\lambda_\nu(E)}\right]
          \equiv Z_\nu^{(\mathrm{I})}(E,x),
\]
where $Z_\nu^{(1)}(E,x)$ is defined by eq.~(\ref{Z0}). Thus
\[
Z_\nu^{(\mathrm{I})}(E,x) = Z_\nu^{(1)}(E,x)\left[1
-\frac{xZ_\nu^{(1)}(E,x)}{2\lambda_\nu(E)}+\ldots\right] \leq 1.
\]
However, the $Z_\nu^{(\mathrm{I})}(E,x)$ can be approximated by
$Z_\nu^{(1)}(E,x)$ with very good accuracy, because
$xZ_\nu^{(1)}(E,x)/\lambda_\nu(E) \ll 1$ in most
cases of interest for neutrino astrophysics.
%%%%%%%%%%%%%%%%%%%%%%%%%%%%%%%%%%%%%%%%%%%%%%%%%%%%%%%%%%%%%%%%
%% But $Z_\nu^{(\mathrm{I})}(E,x) \ll Z_\nu^{(1)}(E,x)$ if
%% $xZ_\nu^{(1)}(E,x)\gg\lambda_\nu(E)$ (unrealistic case). 
%%%%%%%%%%%%%%%%%%%%%%%%%%%%%%%%%%%%%%%%%%%%%%%%%%%%%%%%%%%%%%%%

%\clearpage

\protect\section{Neutrino transport equation with a source function}
\label{app:GF}

Here, we briefly show how to take into account the contributions from
production of neutrinos through reactions
$\nu_\ell T\rightarrow\nu_{\ell'} X$ ($\ell\neq\ell')$ or the
reaction chains mentioned in the introduction in the case when these
may be treated as corrections to the principal solution described in
sect.~\ref{sec:Z} and appendix~\ref{app:MIS}. Clearly, the problem
reduces to the transport equation~(\ref{TE}) with a source function
$S_\nu(E,x)$ on the right side. In line with our general approach,
we will seek the solution to this equation in the following form%
\footnote[7]{We suppose $F_S(E,0)=0$ as the boundary condition.}
\begin{equation}\label{G}
F_S(E,x) = \int_0^x S_\nu(E,x')\exp\left[-\int_{x'}^x
\frac{1-\mathcal{Z}_\nu(E,x'')}{\lambda_\nu(E)}\d x''\right]\d x'
\end{equation}
with $\mathcal{Z}_\nu(E,x)$ a positive function satisfying
the equation
\begin{equation}\label{calZG}
\mathcal{Z}_\nu(E,x) =
               \int_0^1\eta_S(y,E;x)\Phi_\nu(y,E)\d y,
\end{equation}
where we introduced
\[
\eta_S(y,E;x) = \frac{F_S(E_y,x)}{F_S(E,x)(1-y)}.
\]
It is easy to verify that $F_S(E,x) \sim xS_\nu(E,0)$
as $x\rightarrow 0$. Therefore,
\[
\eta_S(y,E;0) = \frac{S_\nu(E_y,0)}{S_\nu(E,0)(1-y)},
\]
and this function is assumed to be finite for any $E$ and $y$.

The algorithm for the solution to eqs.~(\ref{G}), (\ref{calZG})
is quite obvious: putting $\mathcal{Z}_\nu^{(0)} = 0$ yields 
\begin{eqnarray*}
F_S^{(0)}(E,x) & = & \int_0^x S_\nu(E,x-x')
                        \e^{-x'/\lambda_\nu(E)}\d x', \\
\mathcal{Z}_\nu^{(1)}(E,x) & = & \int_0^1\eta_S^{(0)}(y,E;x)
                                 \Phi_\nu(y,E)\d y,
\end{eqnarray*}
etc.
The formal question about the finiteness of the involved integrals
over $y$ is closely related to the very difficult problem of the
asymptotic behavior for the $\nu N$ inclusive and total cross
sections as $E\rightarrow\infty$. This problem is beyond the scope of
this study, but we can avoid it introducing a cutoff $y_0=1-E/E_0$
(with $E_0 \gg E$) as the upper limit of the integrals. The reason
for such a cutoff is in the fact that any physical source function,
$S_\nu(E,x)$, must exponentially vanish as $E\rightarrow\infty$
(cf. footnote~\ref{note:cutoff}).

%\clearpage

\protect\section{Neutrino-nucleon cross sections}\label{app:NNCS}

According to ref.~\cite{Gandhi96}, the inclusive differential cross
sections for the reactions $\nu_\mu N\rightarrow\mu^-X$ (CC) and
$\nu_\mu N\rightarrow\nu_\mu X$ (NC), where $N$ is an isoscalar nucleon,
in the renormalization-group-improved parton model are of the form
($E\gg M_N$)
\begin{eqnarray*}
\frac{\d\sigma_{CC}(y,E)}{\d y} & = & \frac{2G_F^2M_NE}{\pi}\int_0^1
\frac{A(\hat{x},Q^2)+(1-y)^2\overline{B}(\hat{x},Q^2)}
                                     {(Q^2/M_W^2+1)^2}\,\d\hat{x}, \\
\frac{\d\sigma_{NC}(y,E)}{\d y} & = & \frac{G_F^2M_NE}{2\pi}\int_0^1
\frac{A_0(\hat{x},Q^2)+(1-y)^2\overline{B}_0(\hat{x},Q^2)}
                                     {(Q^2/M_Z^2+1)^2}\,\d\hat{x}.
\end{eqnarray*}
Here $G_F$ is the Fermi constant, $M_N$ and $M_W$ ($M_Z$) are the
nucleon and $W$- ($Z$-) boson masses, respectively, $Q^2=2M_N\hat{x}yE$,
is the squared invariant momentum transfer between the incident and 
outgoing lepton, and $\hat{x}$ is the usual Bjorken scaling variable.
For the $\overline{\nu}_\mu N$ cross sections, the structure functions
$A$, $\overline{B}$, $A_0$ and $\overline{B}_0$ in the above formulas
should be substituted for the functions $\overline{A}$, $B$,
$\overline{A}_0$ and $B_0$, respectively. All these are
\begin{eqnarray*}
          A    & = & \frac{1}{2}\left(u_v+d_v+u_s+d_s\right)
                                                    +s_s+b_s,  \\
\overline{A}   & = & \frac{1}{2}\left(u_s+d_s\right)+s_s+b_s,  \\
          B    & = & \frac{1}{2}\left(u_v+d_v+u_s+d_s\right)
                                                    +c_s+t_s,  \\
\overline{B}   & = & \frac{1}{2}\left(u_s+d_s\right)+c_s+t_s,  \\
          A_0  & = & \frac{1}{2}\left(L_u^2+L_d^2\right)
                                \left(u_v+d_v+u_s+d_s\right)
                    +\frac{1}{2}\left(R_u^2+R_d^2\right)
                                        \left(u_s+d_s\right)+C,\\
\overline{A}_0 & = & \frac{1}{2}\left(L_u^2+L_d^2\right)
                                        \left(u_s+d_s\right)
                    +\frac{1}{2}\left(R_u^2+R_d^2\right)
                                \left(u_v+d_v+u_s+d_s\right)+C,\\
          B_0  & = & \frac{1}{2}\left(R_u^2+R_d^2\right)
                                        \left(u_s+d_s\right)
                    +\frac{1}{2}\left(L_u^2+L_d^2\right)
                                \left(u_v+d_v+u_s+d_s\right)+C,\\
\overline{B}_0 & = & \frac{1}{2}\left(R_u^2+R_d^2\right)
                                \left(u_v+d_v+u_s+d_s\right)
                    +\frac{1}{2}\left(L_u^2+L_d^2\right)
                                        \left(u_s+d_s\right)+C,
\end{eqnarray*}
where
\[
C = \left(L_d^2+R_d^2\right)\left(s_s+b_s\right)+
    \left(L_u^2+R_u^2\right)\left(c_s+t_s\right),
\]
$R_d=\frac{2}{3}\sin^2\theta_W$, $R_u=-2R_d$, $L_d=R_d-1$ and
$L_u=R_u+1$ are the chiral couplings, and $\theta_W$ is the
Weinberg angle. The functions $u_{v,s} = u_{v,s}(\hat{x},Q^2)$,
$d_{v,s} = d_{v,s}(\hat{x},Q^2)$, etc. are the distributions of
corresponding valence ($v$) and sea ($s$) quark flavors in a proton.
For all constants involved we used the standard values from
ref.~\cite{PDG96}.

\end{document}